\documentclass[twocolumn,amsmath,amssymb,aps,prl,superscriptaddress]{revtex4-2}
\usepackage{graphicx}
\usepackage{dcolumn}
\usepackage{enumitem}
\usepackage{bm}
\usepackage{hyperref}

\begin{document}
\title{Social contagion with emotional group interactions}

\author{YuQianqian Ma}
\affiliation{School of Science, Beijing University of Posts and Telecommunications, Beijing 100876, China}

\author{Peng Zhang$^*$}
\affiliation{School of Science, Beijing University of Posts and Telecommunications, Beijing 100876, China}

\author{Leyang Xue$^*$}
\affiliation{International Academic Center of Complex Systems, Beijing Normal University, Zhuhai, 519087, China}
\affiliation{School of Systems Science, Beijing Normal University, Beijing, 100875, China}
\email{zhangpeng@bupt.edu.cn, hsuehleyang@gmail.com}

\begin{abstract}
Individual decisions and behaviors are shaped not only by direct interactions with others but also by the collective emotional dynamics within groups.
In this work, we introduce the signed simplicial contagion model,  integrating both pairwise and emotional group interactions to investigate  contagion dynamics in signed networks.
Through mean field analysis and numerical simulations, we show that emotional group interactions can induce discontinuous phase transitions, bistable behavior, and hysteresis loops.
However, as the proportion of negative edges $q$ increases, the influence of group interactions weakens under a given transmission strength, driving a shift from discontinuous to continuous phase transitions. 
Our findings reveal that pairwise and group interactions respond differently to changes in $q$: group interactions display nonlinear sensitivity, while pairwise interactions exhibit a more gradual, linear response.
This divergence shifts the dominant mechanisms of contagion, depending on the levels of trust and distrust in the network, providing deeper insights into how emotional relational shape the spread of contagion in social systems.

\end{abstract}



\maketitle
\section{Introduction}
Social contagion processes, such as disease spread and information diffusion, have become a major focus in network science, with numerous models developed to capture various mechanism and better understand how these dynamics unfold~\cite{latora2017complex, wang2019disease,sun2014epidemic,ferraz2018general,jing2022effective}. 
However, real-world interactions are rarely neutral---emotional tendencies between individuals play a crucial role in shaping these processes~\cite{keltner1998emotion,lopes2004emotional,andersen1996principles}.
For instance, individuals with stronger positive relationships are more likely to interact frequently, potentially increasing the likelihood of transmission compared to those with weaker or negative relationships. 
To model these emotional influences, signed networks, which represent both positive and negative relationships, have proven to be a useful tool in studying social contagion processes~\cite{chengsuqi2013survey,Shi2017DynamicsOS}.

With the introduction of positive and negative emotional interactions into contagion processes, significant progress has been made in several key areas.
From an applied perspective, some studies have focused on designing effective immunization and control strategies in signed networks to either inhibit or promote the spread of information~\cite{li2022immunization,kundu2023rumor}.
On a theoretical level, various contagion models have been developed for signed network, incorporating different real-world scenarios and mechanisms to explore the dynamic processes and uncover the complex phenomena~\cite{liu2019influence,niu2021information,li2021dynamics}.
Additionally, the influence of network structure on dynamic behavior has been extensively examined, with studies investigating how the configuration of signed relationships impacts contagion dynamics and the induced phenomena~\cite{gao2018uncovering,saeedian2017epidemic,zhang2022impact}.

Most existing studies on contagion processes in signed networks have primarily focused on pairwise interaction dynamics, capturing the effects of positive and negative relationships between individuals through direct node-to-node interactions.
However, in social systems, individuals decision-making and behavior are influenced not only by these pairwise interactions but also by the overall emotional climate of the group. 
For instance, in a harmonious group,  individuals are more likely to align with group decisions and adopt shared opinions, 
as a positive group emotional climate fosters trust and openness, making them more comfortable conforming to the collective sentiment~\cite{barsade2002ripple,vanSwol2021FosteringCC}.
In contrast, in a discordant or negative emotional climate, individuals may be less inclined to follow group norms, as the lack of cohesion weakens the group’s influence on individual behavior. 
Despite these mechanisms being widespread in real-world social systems, little is known so far about how group sentiment drives collective contagion processes.

To capture the effect of group emotional interactions on individual decision-making, we propose a novel social contagion model in signed networks.
This model integrates both emotional interactions at the individual level---where positive relationships drive the transmission---and at the group level, where the collective influence is shaped by the emotional cohesion within the group and the consistent adoption of behaviors by its members.
Specifically, in balanced groups (i.e., harmonious or stable), if all members except one have adopted a particular behavior, the group exerts a stronger collective influence on the remaining individual to conform. 
In contrast, in unbalanced groups (i.e., discordant or unstable), this collective influence is absent.

Our model account for both pairwise emotional interactions and group-level emotional dynamics, providing a comprehensive framework for characterizing the interplay between individual and collective emotional behavior in signed networks. 
A key feature of this model is its ability to capture the impact of negative relationships on the contagion process. 
Unlike most models on signed networks, which treat negative relationships as non-interactions~\cite{Guler2014CommunicatingIA,ju2020new}, our model highlights their crucial role in shaping group effects on susceptible nodes.
Instead of overlooking these relationships, our model shows that even small changes in negative interactions can significantly reshape both pairwise and group dynamics, thereby influencing the overall contagion process.
As demonstrated in this paper, the ratio of negative relationship, \textit{q}, significantly alters the dynamic phenomena. 
When \textit{q=0}, our model reduces to the simplicial models typically described in unsigned networks~\cite{iacopini2019simplicial}, further demonstrating its flexibility and general applicability.


\begin{figure}
    \centering
    \includegraphics[width=\linewidth]{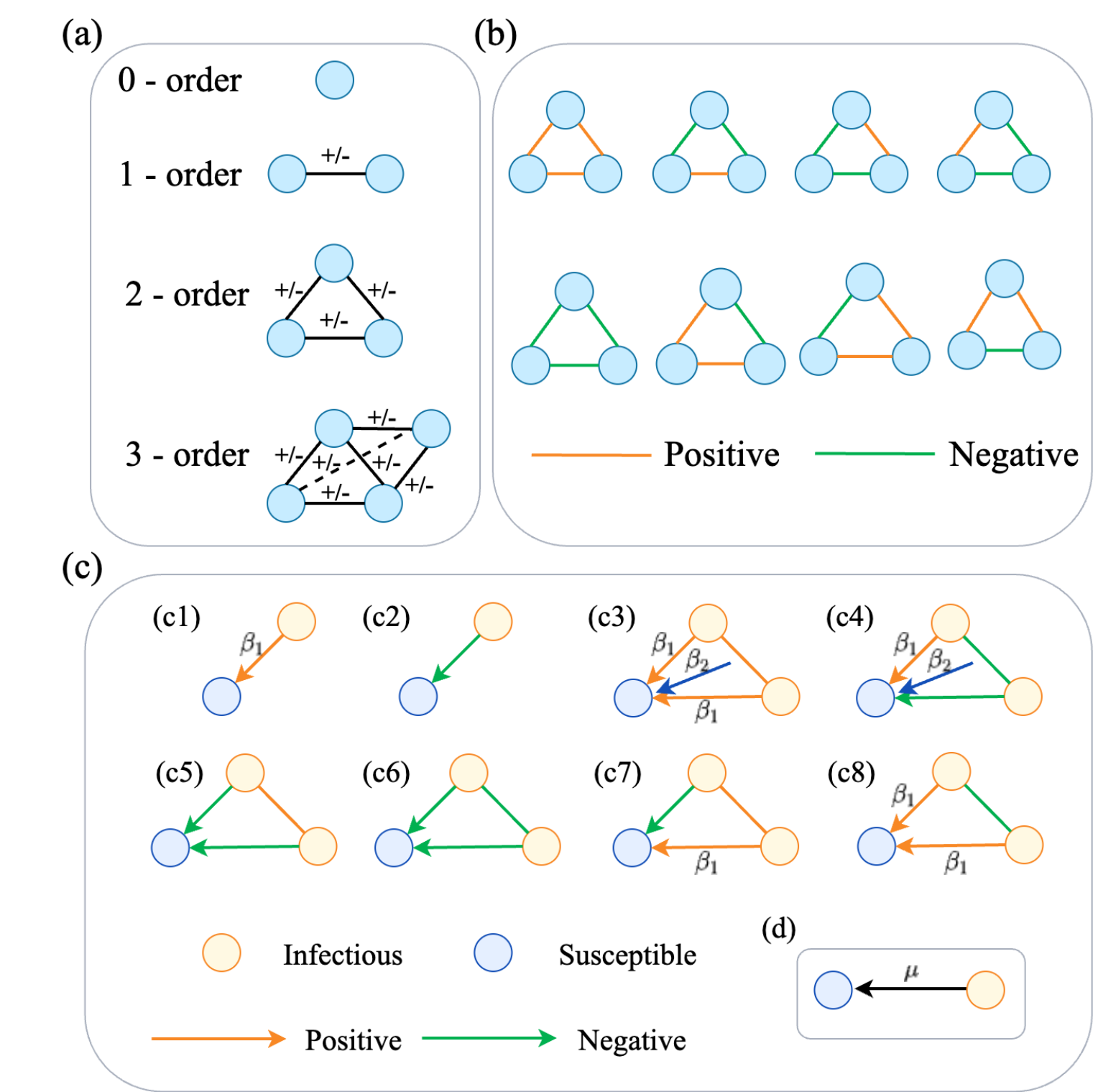}
    \caption{\textbf{Illustration of signed simplicial complexes model~(SSCM)}. 
    \textbf{(a)} Emotional interactions with positive or negative relationships in signed d-simplex (e.g., d=0, 1, 2, 3).
    \textbf{(b)} All possible underlying structures in a signed 2-simplex. 
    The first row shows all balanced triangles, while the second row exhibits all unbalanced triangles.
    \textbf{(c)} Transmission channels for susceptible individuals in signed 2-simplices. 
    In the contagion model, a susceptible node can only be infected through positive links (i.e., interactions with trusted individuals) with a probability of $\beta_1$. 
    No infection occurs through negative links. 
    For group interactions involving emotions, transmission occur with a probability of $\beta_2$ only if the interactions are balanced and the consistency condition holds (i.e., all members except one have adopted a particular behavior); otherwise, such effects do not occur.
    As described in \textbf{(c1)-(c8)}, only the balanced triangular structures shown in (c3) and (c4) can induce emotional group interaction effects on the susceptible nodes.
    \textbf{(d)} All infected nodes spontaneously recover to the susceptible state at a rate of $\mu$.
    }
    \label{fig:1}
\end{figure}

In this paper, we introduce a signed simplicial contagion model~(SSCM) to study how emotional group interactions and negative relationships shape contagion dynamics in signed networks. 
Our findings demonstrate that emotional group interactions can trigger discontinuous phase transitions, leading to bistable states and hysteresis loops. 
However, as the proportion of negative edges $q$ increases, the strength of group interactions diminishes, resulting in the gradual disappearance of these discontinuous transitions and a shift toward continuous contagion dynamics.

We further demonstrate that increasing $q$ not only alters the nature of phase transitions but also shifts their critical thresholds. 
A key insight from our study is that pairwise interactions and group interactions respond differently to changes in the proportion of negative edges $q$.
While group interactions weaken nonlinearly with rising $q$, pairwise interactions exhibit a more direct, linear dependence on the same parameter. 
This asymmetry sheds light on how distrustful relationships reduce the influence of both interaction types, transforming the contagion process in distinct ways.

Our results, supported by both numerical simulations and mean field analysis, highlight the critical role of emotional group interactions in shaping contagion dynamics.
These insights offer valuable guidance for managing social dynamics---such as misinformation, panic, or unrest---in real-world networks, where the balance between positive and negative relationships plays a pivotal role in determining contagion outcomes.

The remainder of the paper is organized as follows:
In Section 2, we describe the dynamics of social contagion, covering the contagion model, contagion process, and Section 3 gives the formation of the simplicial complexes with signs and an analytical analysis using mean field theory.
Section 4 presents the numerical simulation results of the dynamics and demonstrates the effect of the ratio of negative relationships on contagion.
Finally, Section 5 provides a summary and discussion of our findings.

\section{Model}
\textbf{The contagion model}.
In real-world social interactions, individuals are more likely to be influenced by those with whom they share a strong positive connection than by those with whom they have conflicts.
Similarly, when individuals belong to a harmonious group where members exhibit consistent behavior, they are more susceptible to the group’s collective influence~\cite{li2022structural, heider1946attitudes}.
Each type of interaction---pairwise or group---independently shapes adoption behavior through different mechanisms: individual relationships foster trust and direct influence, while group dynamics create collective pressure for conformity. 
Our model capture these key interactions that are central to the contagion process, with transmission occurring in two distinct pathways: pairwise interactions driven by trusted relationships and group interactions where group harmony promotes uniform behavior adoption.

To formalize these dynamics, we utilize signed simplicial complexes---mathematical structures that extend traditional network representations by incorporating higher-order interactions.
A signed $k$-simplex represents a group of $k+1$ individuals, $V=\{v_0, v_1, \dots ,v_k\}$, where each pair is connected by a link that carries either a positive or negative sign: a positive sign~(+1) indicates a friendly or cooperative relationship, while a negative sign~(–1) represents an antagonistic or conflicting relationship.
By assigning these signs to the links, we form simplices of varying dimensions that capture group dynamics beyond what pairwise relationships alone can describe (see Fig.~\ref{fig:1}(a)).
This framework allows us to model both the nature of individual relationships and the overall emotional balance within groups.
The signs on the links of a simplex illustrate how individual connections contribute to either group harmony or discord.

To evaluate group harmony, we incorporate structural balance theory, originally proposed by Heider~\cite{heider1946attitudes}. 
This theory provides a foundation for understanding the stability and cohesiveness of social groups based on the configuration of positive and negative relationships among members.
The core principles are: a friend of a friend is a friend, an enemy of an enemy is a friend, a friend of an enemy is an enemy, and an enemy of a friend is an enemy.
Based on these principles, we classify groups as either balanced, where the product of the signs of all relationships is positive, or unbalanced, where the product is negative.
Balanced groups are harmonious, fostering strong collective influence that promotes conformity and facilitates the contagion.
In contract, unbalanced groups contain conflicting relationships, leading to instability and reducing the effectiveness of group influence on individual behavior.

For example, in a 2-simplex (a triangle of three individuals), where each edge is assigned a random positive or negative sign, there are eight possible combinations of signs, as shown in Fig.~\ref{fig:1}(b).
Applying structural balance theory, we categorize these triangles into balanced and unbalanced configurations: the first row illustrates balanced group, while the second row shows unbalanced ones.
These classification extends naturally to groups of any size.
By integrating structural balance theory with signed simplices, our model captures how both individual relationships and group harmony shape the contagion process.

\textbf{The contagion process}.
Our proposed model, the Signed Simplicial Contagion Model (SSCM), is designed to capture contagion dynamics in signed networks by incorporating both pairwise and group-level interactions. 
In the SSCM, each node can exist in one of two states: susceptible ($S$) or infectious ($I$), represented as $x_i(t)\in \{0,1\}$, where $0$ indicates a susceptible state and $1$ represents an infectious state.

The transmission mechanism between states depends on the relationships between nodes.
For simplicity, we consider interactions up to the 2-simplex, with a set of infection parameters $B=\{\beta_1,\beta_2\}$.
In pairwise interactions, a susceptible node can be infected by an infectious neighbor through a positive edge, with probability $\beta_1$~($S+I \rightarrow 2I$), whereas no infection occurs across negative edges.
At the group level, infection spreads within balanced triangle structures (2-simplices) according to the parameter $\beta_2$, requiring at least one positive edge between an infectious and a susceptible node for higher-order contagion to take place.
For example, if a susceptible individual is connected to two infectious nodes by negative edges, the group interaction does not contribute to contagion in this case. 

Figure~\ref{fig:1}(c), panels (c1)–(c8), illustrate this transmission mechanism, highlighting two cases where group interactions ($\beta_2$) facilitate contagion. 
The model also incorporates a recovery mechanism, where infectious nodes transition back to the susceptible state with probability $\mu$ ($I \rightarrow S$), as shown in Fig.~\ref{fig:1}(d).

The system’s contagion process is tracked by the order parameter $\rho(t)=\frac{1}{N}\sum_{i=1}^{N}x_i(t)$, representing the density of infected nodes at time $t$.
In each numerical simulation, we record the value of $\rho(t)$ at the every time step. Since $\rho(t)$ evolves gradually, the value at the final time step may not accurately reflect the overall result of the simulation.
To ensure reliable results, we run the simulation for a sufficiently long period, setting $t = 6000$ to allow the system to reach a relatively steady state. 
The final infection density, $\rho^*$, is computed by averaging $\rho(t)$ over the last 100 time steps, i.e., $\rho^* = \frac{1}{100}\sum_{t=5901}^{6000}\rho(t)$.
This value represents the outcome of one simulation.
For a given $\beta_1$, the final result is the average of 120 independent simulation outcomes. 

\section{Method}
\textbf{Random signed simplicial complexes}. 
We describe the procedure for generating signed simplices in this work. 
Starting with the parameters $\langle k \rangle$ and $\langle k_{\triangle} \rangle$, which represent the average degree and the average number of triangles per node, respectively, 
we first construct an Erd\H{o}s-R\'enyi~(ER) network with $N$ nodes. 
In this network, any two nodes $i$ and $j$ are connected with a probability $p_1$.
Following this, we randomly form 2-simplices by connecting any combination of three nodes $(i, j, k)$ with a probability $p_2$. 
These probabilities, $p_1$ and $p_2$, are expressed in terms of the parameters $p_1=\frac{\langle k \rangle-2\langle k_{\triangle} \rangle}{(N-1)-2\langle k_{\triangle} \rangle}$, and $p_2 = \frac{2\langle k_{\triangle} \rangle}{(N-1)(N-2)}  $~\cite{iacopini2019simplicial}.
Next, we introduce the parameter $q$ to represent the proportion of negative edges in the network. 
For a given $q$, a fraction $q$ of the edges, selected at random, are assigned as negative. 
Once the edge signs are assigned, the network forms a synthetic signed simplicial complex, with an average positive degree of $\langle k_+ \rangle = \langle k \rangle(1 - q)$ and an average negative degree of $\langle k_- \rangle = \langle k \rangle q$.

Similar to the positive degree, we define $\langle k_{B} \rangle$ as the average number of balanced triangles connected to each node, representing a potential channel for group interaction.
This is given by:
\begin{equation}\label{Eqs:eq1}
\langle k_{B} \rangle=P_{B}\langle k_{\triangle} \rangle, 
\end{equation}
where $P_{B}$ denotes the probability that a triangle is balanced. 
This probability depends on the proportion of negative edges $q$, and is given by the following equation:
\begin{align}\label{Eqs:eq2}
    P_{B}&=(1-\textit{q})^3+C_3^2\textit{q}^2(1-\textit{q})  \nonumber \\
    &=-4\textit{q}^3+6\textit{q}^2-3\textit{q}+1.
\end{align}
This formulation captures the interplay between the proportion of negative edges and the balance within triangles, highlighting how these factors influence the average number of balanced triangles connected to each node.
To validate this theoretical description, we count balanced triangles in the generated signed network and compare the results with the theoretical values derived from Eq.~\ref{Eqs:eq1}.
This comparison demonstrates the accuracy of derived expression, as illustrated in Fig.~\ref{fig：numerical_theory} in supplementary information.

\textbf{Mean field approach.} 
To accurately track the dynamical behavior of contagion processes, we employ a mean field approach for analytical description~\cite{su2022analysis,sahneh2013generalized,jones2023improving}.
For any given node, the probability of remaining in the infected state at time $t$ is denoted as $\rho(t)$, representing the temporal evolution of the density of infected nodes.
This probability follows the differential equation:
\begin{align}\label{Eqs:eq3}
{\rm d}_t\rho(t)=-\mu\rho(t)+\beta_1\langle k_+\rangle \rho(t)[1-\rho(t)]\notag\
\\+\beta_2\langle k_{B}\rangle \rho^2(t)[1-\rho(t)],
\end{align}
where $\mu$ is the recovery rate, $\beta_1$ and $\beta_2$ are the transmission probability associated with pairwise interactions~(positive edges) and emotional group interactions (balanced triangles), respectively.
The first term $-\mu\rho(t)$ accounts for the recovery of infected nodes.
The second term reflects the contribution of pairwise interactions~(positive edges), while the third term captures the influence of group interactions~(balanced triangles) on the infection dynamics.

To obtain a general result independent of specific system parameters, we rescale the transmission probabilities~($\beta_1$, $\beta_2$) based on the network structure and recovery rate.
Specifically, we define:
\begin{equation}\label{Eqs:rescaling}
\lambda_1=\frac{\beta_1\langle k_+ \rangle}{\mu}, \lambda_2=\frac{\beta_2\langle k_{B} \rangle}{\mu}.
\end{equation}
By rescaling the time with the recovery rate $\mu$, we can rewrite Eq.~\ref{Eqs:eq3} as:
\begin{equation}\label{Eqs:eq4}
{\rm d}_t\rho(t)=-\rho(t)[1-\lambda_1+(\lambda_1-\lambda_2)\rho(t)+\lambda_2\rho^2(t)].
\end{equation}
This rescaling simplifies the analysis by reducing  the dependence on specific parameters while still capturing the essential dynamics of the contagion process.

When the system reaches a steady state~(i.e., $t \rightarrow \infty$, denoted as $\rho^*$), the rate of change becomes zero, allowing us to evaluate the equilibrium conditions and analyze the system’s stability~\cite{strogatz2018nonlinear}. 
This leads to the equation ${\rm d}_t \rho^* = 0$, from which we can determine the equilibrium values of $\rho^*$, yielding the following solutions:
\begin{equation}\label{Eqs:eq5}
\rho_{2_{\pm}}^*=\frac{\lambda_2-\lambda_1\pm\sqrt{(\lambda_1-\lambda_2)^2-4\lambda_2(1-\lambda_1)}}{2\lambda_2},
\end{equation} 
where $\rho_1^* = 0$ is one of the solutions, corresponding to the typical absorbing, epidemic-free state in which all individuals have recovered.
The two other roots, $\rho^*_{2_{\pm}}$, represent the potential equilibrium densities of infected nodes, depending on the system parameters. 
Further stability analysis of these solutions helps determine the conditions under which the system exhibits either stable or unstable dynamics.

\begin{figure}
    \centering
    \includegraphics[width=1\linewidth]{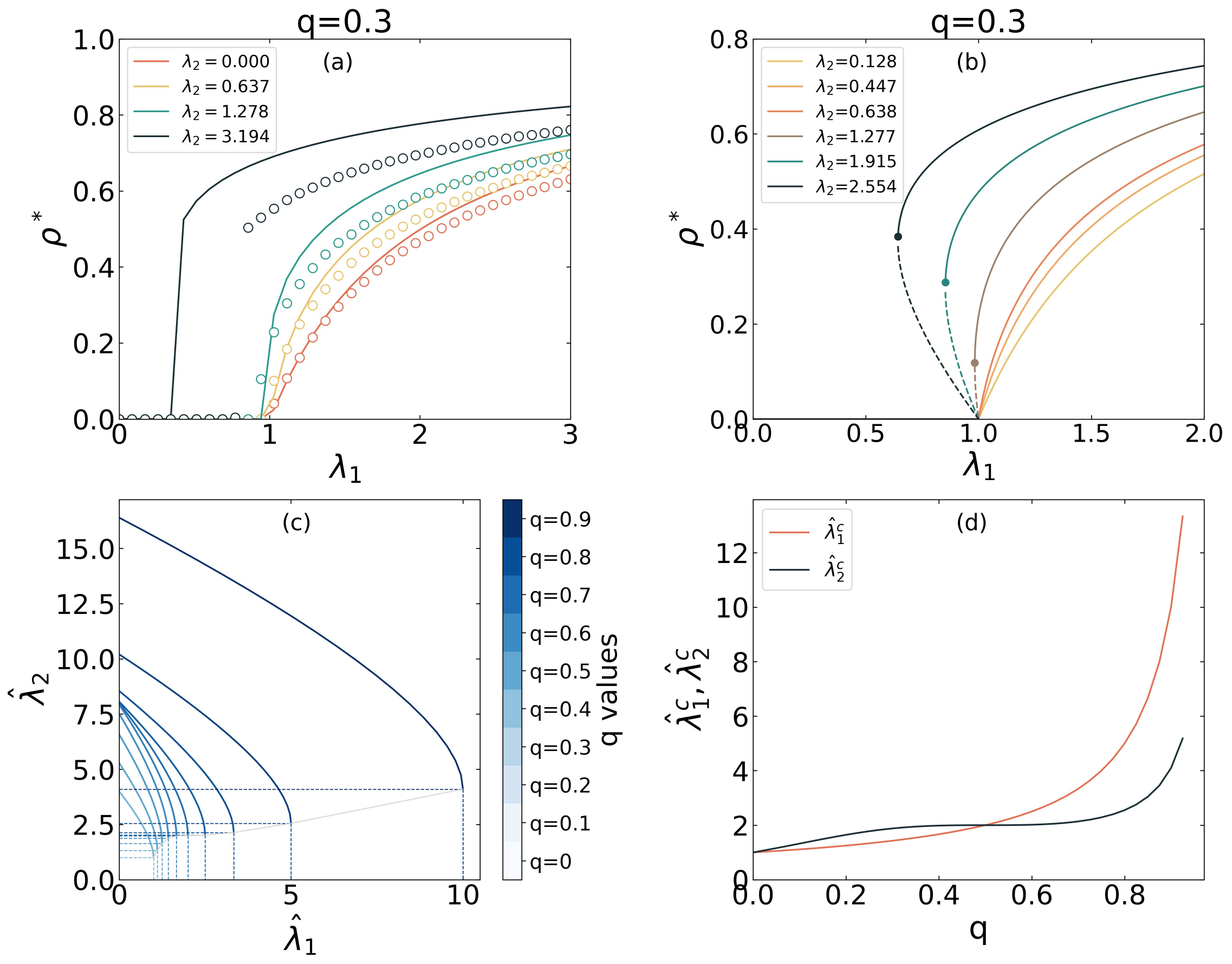}
    \caption{\textbf{Analytical results derived from the mean field method}.
    \textbf{(a)} Comparison of the mean field method results~(lines) with simulations~(circles), demonstrating the accuracy of the mean field approach.  
    \textbf{(b)} The stationary solutions $\rho^*$ for different values of $\lambda_2$ are plotted as a function of $\lambda_1$, as given by Eq.~\ref{Eqs:eq5}.  
    Dashed and solid lines represent unstable and stable solutions, respectively.
    Unstable solutions exist only when $\lambda_2 > 1$.
    \textbf{(c)} Plot of $\hat{\lambda}_2(q)$ versus $\hat{\lambda}_1(q)$ for different values of $q$, where $\hat{\lambda}_2(q)=\frac{\lambda_2}{P_B(q)}$ and $\hat{\lambda}_1(q) = \frac{\lambda_1}{1 - q}$.
    These parameters incorporate the influence of the proportion of negative edges, illustrating how the critical behavior changes with varying $q$.
    For each given value of $q$, the region enclosed by the corresponding curve and the lower left-hand corner represents the area where $\rho^* = 0$, indicating no infection. 
    The region on the opposite side corresponds to $\rho^* \neq 0$, where infection persists.
    The vertical dashed lines for each $q$ mark the critical threshold $\hat{\lambda}^c_1$ of a standard SIS contagion process without group interaction effects,
    while the horizontal dashed lines indicate the threshold of group interactions $\hat{\lambda}^c_2$ that trigger a discontinuous transition.
    \textbf{(d)} The effect of the proportion of negative edges in the system on the critical thresholds $\lambda_1^c$ and $\lambda_2^c$.
    }
    \label{fig:enter-label}
\end{figure}

To analyze the fixed points in Eq.~\ref{Eqs:eq5}, we examine the discriminant: 
\begin{equation}
D = (\lambda_1 - \lambda_2)^2 - 4\lambda_2(1-\lambda_1).
\end{equation}
The fixed points $\rho^*_{2_{\pm}}$ are real only when $D\geq 0$.
Given the significant influence of group interactions on the contagion dynamics, we analyze two distinct cases: (1) $\lambda_2 \leq 1$ and (2) $\lambda_2 > 1$. 

Case 1: $\lambda_2 \leq 1$.
If $\lambda_1 \leq 1$, the only valid fixed point is $\rho^*=0$, which is stable, indicating no sustained infection.
if $\lambda_1 > 1$, since $D>0$, a positive, stable fixed point $\rho^*_{2+}>0$ emerges, and $\rho^*=0$ become unstable.
This marks the onset of a stable epidemic state.

Case 2: $\lambda_2 > 1$.
To determine the critical infection threshold $\lambda_c$, we set the discriminant $D=0$ and solve the equation $(\lambda_1 - \lambda_2)^2 - 4\lambda_2(1-\lambda_1) = 0$.
Expanding and rearranging this yields: 
\begin{equation}\label{Eqs:lambdac}
\lambda_c = 2\sqrt{\lambda_2} - \lambda_2.
\end{equation}
Note that $\lambda_c \in [0,1)$, because $\lambda_c$ decreases monotonically as $\lambda_2$ increases for $\lambda_2 > 1$.
If $\lambda_1 < \lambda_c$,  since $D<0$, the fixed points $\rho^*_{2{\pm}}$ are complex, and the only stable solution is $\rho^*= 0$ (no infection);
If $\lambda_c \leq \lambda_1 < 1$, in this range, $D \geq 0$, and three real fixed points exist: $\rho^*= 0$ and $\rho^*_{2_{\pm}}$, with $0 < \rho^*_{2-} < \rho^*_{2+}$.
Both $\rho^* = 0$ and $\rho^*_{2+}$ are stable, while $\rho^*_{2-}$ is unstable.
This bistable behavior means the final outcome depends on the initial infection density. 
If the initial density is below $\rho^*_{2-}$, the system settles into the absorbing state ($\rho^* = 0$). 
Otherwise, it shifts toward the epidemic state.
The phase transition is discontinuous in this range because $\rho^*_{2+}>0$, indicating a sudden jump in the infection density.
If $\lambda_1 \geq 1$, here, $D\geq 0$, and the fixed points satisfy $\rho^*_{2-} < 0 < \rho^*_{2+}$.
In this case, $\rho^*_{2+}$ is the stable fixed point, while $\rho^*=0$ becomes unstable.

We present the results of the mean field method alongside the simulation outcomes in Fig.~\ref{fig:enter-label}(a). 
With a given proportion of negative edges, $q=0.3$, in the SSCM model, the analytical results~(lines) align well with the simulation results~(circles) for smaller values of $\lambda_2$, confirming the accuracy of analytical predictions.
However, as $\lambda_2$ increases, the agreement between the two begins to deteriorate.
This is due to the system's growing sensitivity to local structural variations, such as correlations between nodes or triangles, which are not fully captured by the mean field approach.

Although the mean field approach shows some discrepancies, it provides key insights into the dynamics of contagion processes involving emotional group interaction effects.
As seen in Fig.~\ref{fig:enter-label}(b), the system displays rich dynamical behavior as parameters vary.
For $\lambda_2>1$, the dashed lines representing $\rho^*_{2-}$ in Eq.~\ref{Eqs:eq5} are unstable, dividing the state space into two distinct regions, indicating bistable behavior.
If the initial infection density exceeds $\rho^*_{2-}$, the contagion dynamic evolve into an endemic state; otherwise, they converge toward the endemic-free state.
Within this parameter range~(i.e., $\lambda_2 > 1$), $\rho^*$ undergoes an abrupt jump at the critical threshold $\lambda_1 = \lambda_c$ or $\lambda_1 = 1$, signaling a discontinuous transitions.
As discussed later, this behavior leads to the formation of a hysteresis loop.

To systematically understand how pairwise interactions and emotional group interaction strength influence the system’s dynamics, we present phase diagrams spanned by $\hat{\lambda}_1(q)$ and $\hat{\lambda}_2(q)$ for different proportions of negative edges, $q$, as shown in Fig.~\ref{fig:enter-label}(c).
We define the relations $\hat{\lambda}_2(q) = \frac{\lambda_2}{P_B(q)}$ and $\hat{\lambda}_1(q) = \frac{\lambda_1}{1 - q}$ to account for the impact of negative relationships on the contagion process, revealing how they shape the dynamics with emotional group interactions.
When $q = 0$, $\hat{\lambda}_2(q) = \lambda_2$ and $\hat{\lambda}_1(q) = \lambda_1$, reducing the model to the higher-order contagion scenario in unsigned networks~\cite{iacopini2019simplicial}.
Each colored curve in Fig.~\ref{fig:enter-label}(c) includes: (1) a vertical dashed line representing the epidemic threshold in the standard SIS model, denoted as $\hat{\lambda}^c_1$, and (2) a horizontal dashed line indicating the critical threshold of emotional group interaction required to trigger a discontinuous phase transition, denoted as $\hat{\lambda}^c_2$.
The solid curves illustrate how varying emotional group interaction strength affects the critical point $\lambda_c$ for pairwise interactions. 
Notably, $\lambda_c$ differs from $\hat{\lambda}^c_1$, since it represents the threshold for pairwise interactions that induce a discontinuous transition driven by group emotional effects.

When $q = 0$, both $\hat{\lambda}_1^c = 1$ and $\hat{\lambda}_2^c = 1$, indicating that the effects of negative relationships have been fully rescaled out.
However, once group emotional interactions are introduced, the impact of negative relationships on the system's critical behavior becomes nontrivial.
As shown by the gray line in Fig.~\ref{fig:enter-label}(c), increasing the proportion of negative relationships significantly raises both $\hat{\lambda}_1^c$ and $\hat{\lambda}_2^c$, indicating that stronger negative relationships make it more difficult to trigger both continuous and discontinuous phase transitions.
Notably, the impact of negative relationships differs between pairwise and group interactions and follows a nonlinear pattern.
For small values of $q$, changes in $q$ have a relatively minor effect on $\hat{\lambda}_1^c$.
However, as $q$ increases approximately 0.5, the sensitivity of pairwise interactions to the proportion of negative edges increases, while the influence on emotional group interactions remains limited (see Fig.~\ref{fig:enter-label}(d)). 
This shift reflects a transition in the dominant mechanisms driving the contagion process:
when negative edges are sparse, pairwise interactions dominate the dynamics. 
In contrast, as the proportion of negative edges grows, group interactions become the primary factor shaping the system’s behavior.

Our findings suggest that in systems with a high proportion of negative relationships, group dynamics~(emotional group interactions) dominate the contagion process, resulting in a discontinuous phase transition. 
In contrast, in systems with fewer negative edges, individual relationships (pairwise interactions) play a more significant role in driving contagion, leading to a continuous transition.
These insights are crucial for designing targeted strategies to manage the contagion based on the network’s relationship characteristics.
For instance, in social networks with many negative edges, sudden outbreaks of behaviors---such as misinformation, panic, or unrest---become more likely due to the non-linear nature of group-driven contagion.

\section{Result}
In the previous section, we employed the mean field approach to derive theoretical descriptions and present analytical analysis.
Although some subtle differences exist between the numerical and theoretical results, the equilibrium-state solutions and their stability offer valuable insights into the critical points and the nature of phase transitions in contagion dynamics.
A key finding is that the proportion of negative edges significantly influences the critical thresholds for pairwise and group interactions, but in fundamentally different ways. 
This distinction shifts the dominant mechanisms driving the contagion process under varying conditions, providing a deeper understanding of how negative edges shape the contagion dynamics.

The mean field approach has limitations, as it neglects certain correlations between higher-order structures.
To overcome these limitations and further investigate the behavior of contagion dynamics, we conduct numerical simulations using the SSCM.
These simulations allow us to test whether the theoretical insights hold under more realistic conditions and capture the finer effects of emotional group interactions in networks with varying proportions of negative edges.

\begin{figure}
    \centering
    \includegraphics[width=\linewidth]{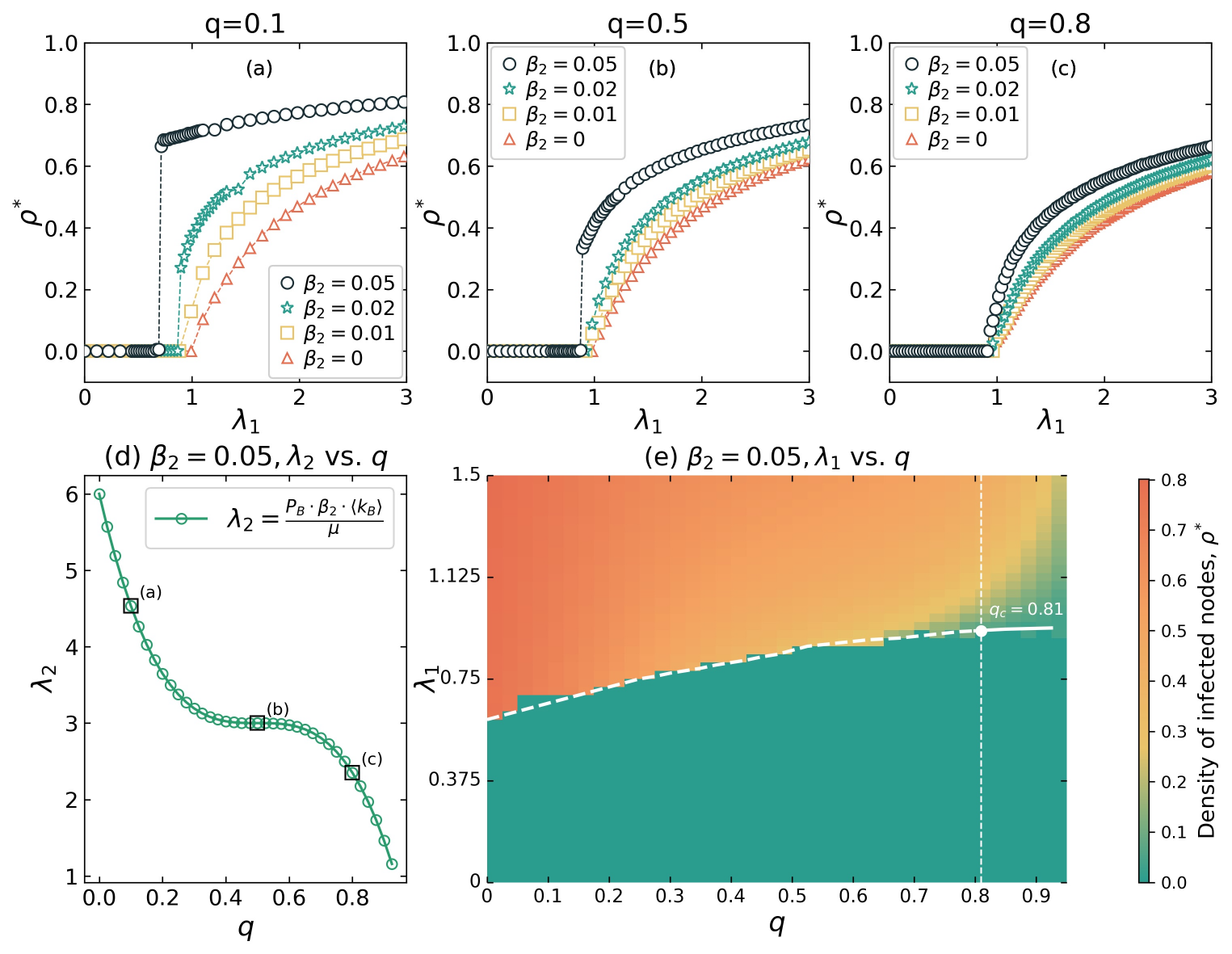}
    \caption{\textbf{Effect of negative relationship on the phase transition of SSCM in synthetic random signed simplicial complexes}.
    We generate RSSC with configuration parameters $\langle k \rangle =20$, $\langle k_{\Delta} \rangle =6$, and $N=2000$.
    \textbf{(a)-(c)} Average density of infectious nodes $\rho^{*}$ as a function of the rescaled pairwise infection probability $\lambda_1$ for different group interaction infection probabilities $\beta_2$. 
    In (a), with a lower proportion of negative edges $p$, increasing $\beta_2$ (e.g., from $\beta_2 = 0$ to $\beta_2 = 0.05$) shifts the transition from continuous to discontinuous.  
    Notably, $\beta_2 = 0$ represents the standard SIS model without the effects of emotional group interactions.
    \textbf{(d)} The relationship between the rescaled group interaction transmission probability, $\lambda_2$, and $q$, with $\beta_2 = 0.5$.
    The values of $q$ used in panels (a)–(c) are marked with black squares.
    \textbf{(e)} Phase diagram of $\rho^{*}$ as a function of \textit{q} and $\lambda_1$.
    The white line denotes the critical infection probability, dividing the phase space into two regions: (i) an absorbing state with no infected individuals, and (ii) an endemic state where a finite proportion of nodes is infected.
    Notably, there is a critical threshold for the ratio of negative edges, denoted as $q_c$. 
    To the left of this threshold (solid line), the SSCM exhibits a discontinuous phase transition, whereas to the right (dashed line), the transition becomes continuous.
    All results are averaged over 120 independent simulations.
    }
    \label{fig:3}
\end{figure}

\textbf{The impact of the proportion of negative relationships.}
In our simulations, we generate signed networks with $N = 2000$, $\langle k \rangle = 20$, $\langle k_{\triangle} \rangle = 6$, and $\mu = 0.05$.
The focus of these simulations is to investigate how varying proportions of negative edges affect the contagion process under a given group interaction strength.
Specifically, we vary the proportion of negative edges, $q$, across three values: $q = \{0.1, 0.5, 0.8\}$, with an initial infection density of $\rho(0) = 0.01$.
For each value of $q$, we test different group interaction infection probabilities, $\beta_2 = \{0, 0.01, 0.02, 0.05\}$. 
The case $\beta_2 = 0$ corresponds to the ordinary SIS model with only pairwise transmission, serving as a baseline for comparison with scenarios involving group interactions. 
We average the final infected node density $\rho^*$ over multiple numerical simulations and present the results in Fig.~\ref{fig:3}(a)-(c), with each panel corresponds to a different values of $q$.
The four symbols in each graph represent varying $\beta_2$ values, allowing us to examine how group interactions influence phase transitions under different proportions of negative edges.

The results show that for all values of $q$, the curves exhibit a continuous transition when $\beta_2 = 0$, reflecting the behavior of the standard SIS model.
However, in Fig.~\ref{fig:3}(a)~($q = 0.1$), as $\beta_2$ increases to $0.05$, $\rho^*$ at the critical point $\lambda_1^c$ exhibits an abrupt jump, indicating a discontinuous phase transition.
This suggest that the introduction of emotional group interaction effects can trigger sudden outbreaks, making the system more prone to rapid shifts between disease-free and endemic states.
As $q$ increases, this discontinuous behavior weakens. 
For instance, in Figs~.\ref{fig:3}(b)($q = 0.5$) and (c)($q = 0.8$), the discontinuous transition becomes less pronounced, particularly for $\beta_2 = 0.02$.
As discussed later, changes in $q$ significantly affect the strength of emotional group interactions, thereby weakening the discontinuous transitions.
This phenomenon is also observed in the mean field analytical results shown in Fig.~\ref{fig:enter-label}(b), where, at $\lambda_1 = 1$, the magnitude of the discontinuous jump decreases as $\lambda_2$ is reduced.
Additionally, the position of the phase transition shifts with increasing $q$.
This result differs from the mean field approximation, where the critical point occurs at $\lambda_1=1$ when $\rho(0) < \rho^*_{2-}$. 
The discrepancy likely arises from correlations between higher-order structures that deviate from the assumptions of mean field theory. 
These findings demonstrate that the proportion of negative edges not only affects the nature of the phase transition but also shifts the position of the critical point, even with fixed group interaction strength.

The previous analysis shows that emotional group interaction effects can trigger discontinuous transitions in contagion dynamics on signed networks.
However, this behavior depends on both the value of $\beta_2$ and the proportion of negative edges, $q$. 
To gain deeper insights, we examine the relationship between $q$ and the rescaled group interaction transmission probability, $\lambda_2 = \frac{P_B \cdot \beta_2 \cdot \langle k_{\triangle} \rangle}{\mu}$, as shown in Fig.~\ref{fig:3}(d). 
We fix $\beta_2 = 0.05$ to demonstrate how $\lambda_2$ varies with different proportions of negative edges.

The results show that as $q$ increases, $\lambda_2$ decreases, indicating that group interaction effects weaken with more negative edges. 
This weakening explains the gradual fading of discontinuous transitions seen in Figs.~\ref{fig:3}(a)–(c), even with the same $\beta_2 = 0.05$. 
Interestingly, $\lambda_2$ exhibits a nonlinear dependence on $q$, in contrast to the linear relationship seen in pairwise interactions.
Specifically, $\lambda_2$ drops sharply at small $q$, stabilizes into a plateau at intermediate values, and declines more slowly for $q > 0.6$. 
This pattern indicates that group interaction dynamics are more sensitive to small changes in $q$ but stabilize as $q$ approaches intermediate levels.

We demonstrate the dependence of $\lambda_2$ on $q$ for a given $\beta_2$.
Now, we turn our attention to the relationship between $\lambda_1$ and $q$.
Using the same $\beta_2$ as in Fig.~\ref{fig:3}(c), we present a heatmap of the final infected node density as a function of $\lambda_1$ and $q$ in Fig.~\ref{fig:3}(e).
In this figure, we identify the critical thresholds based on numerical simulations to illustrate the relationship between $\lambda_1^c$ and $q$, as highlighted by the white line.
The results align with the findings from Figs.~\ref{fig:3}(a)–(c), showing that as $q$ increases, the transition threshold of $\lambda^c_1$ rise. 
This deviation from the mean field analytical result, $\lambda^c_1 = 1$ when $\rho(0) < \rho^*_{2-}$, is possibly attributed to the limitations of the mean field approach, which fails to account for higher-order correlations between nodes.
At smaller values of $q$, the heatmap reveals a sharp color shift from darker to lighter shades as $\lambda_1$ increases, indicating a discontinuous transition~(see the dashed line). 
This occurs because, with fewer negative edges, the emotional group interaction effect remains strong,  leading to a large $\lambda_2$.
In contrast, for larger values of $q$, the group interaction effect weakens, resulting in a more gradual color transition that reflects a continuous transition~(see the solid line).
Hence, a critical proportion of negative edges exists where the phase transition shifts from discontinuous to continuous, marked by the vertical dashed line.
This further demonstrates increasing $q$ influences the nature of the phase transition for a given group emotional interaction strength.
 
\textbf{The bistable phenomenon and hysteresis loops.}
Previous mean field analyses have demonstrated that the initial contagion density, $\rho(0)$, plays a crucial role in determining the final propagation outcomes.
In this section, we perform numerical simulations to investigate how different values of $\rho(0)$ affect the dynamic behavior of our model.

\begin{figure}
    \centering
    \includegraphics[width=1\linewidth]{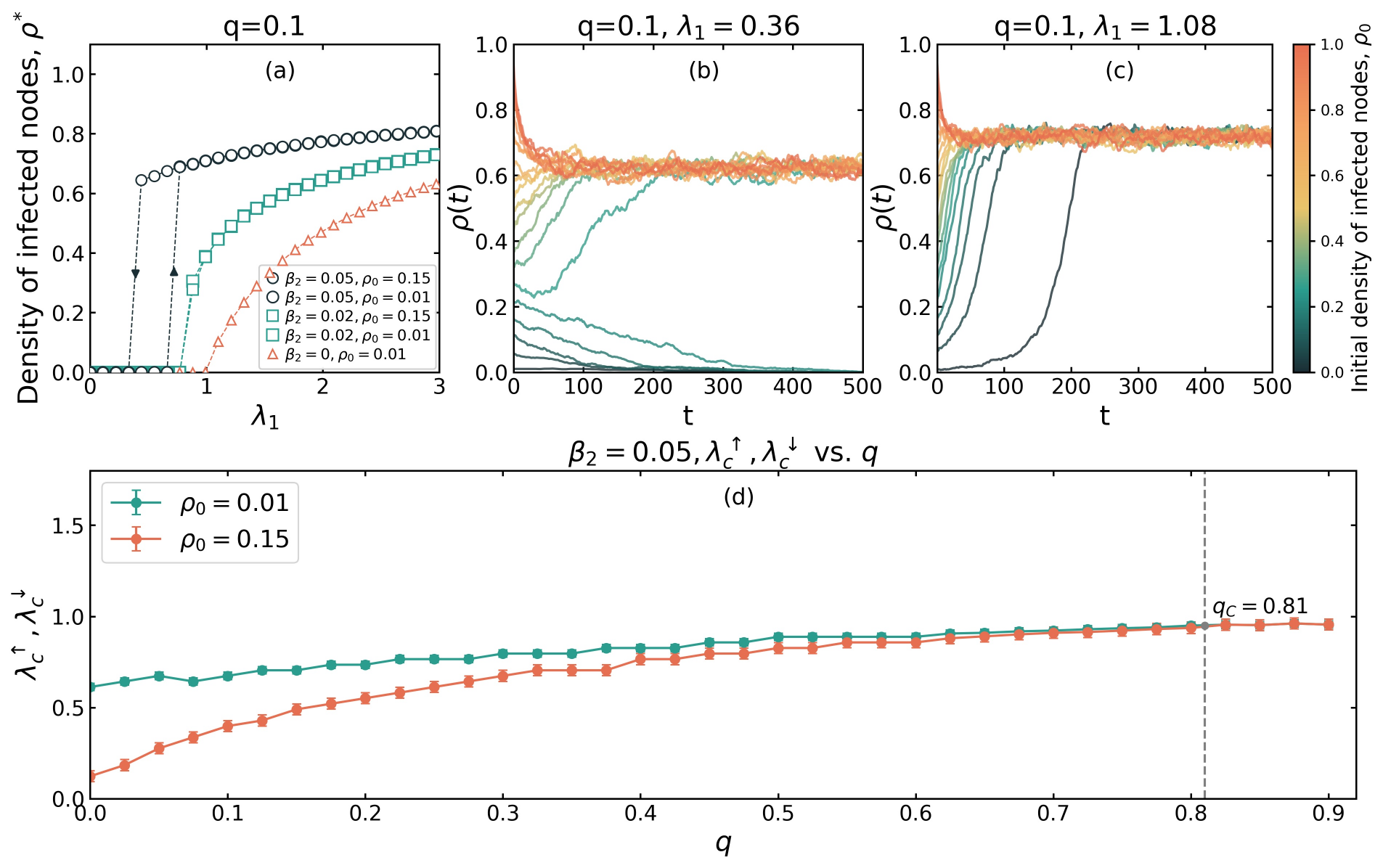}
    \caption{\textbf{Bistable reigion and the effect of proportion of negative relationship $q$.}
    \textbf{(a)} The hysteresis loop. 
    The average final densities of infected nodes are plotted as a function of $\lambda_1$ for different initial densities of infected nodes, $\rho(0)$.
    \textbf{(b)-(c)} The dynamic evolution of the infected nodes densities, $\rho(t)$, initiated from different $\rho(0)$ values, with the same $q$ but varying $\lambda_1$.
    The colors represent the different initial densities of infected nodes in the signed network. 
    Bistable phenomenon are present in (b) but absent in (c).
    \textbf{(d)} The dependence of the critical thresholds, $\lambda_c^{\uparrow}$ and $\lambda_c^{\downarrow}$, identified by numerical simulation, on the proportion of negative edges $q$.
    The critical thresholds $\lambda_c^{\uparrow}$ and $\lambda_c^{\downarrow}$ represent the contagion onset for initial infection densities of $\rho(0)=0.01$ and $\rho(0)=0.15$, respectively.
    }
    \label{fig:4}
\end{figure}

Given parameters $q$, $\beta_1$, and $\beta_2$, we present the final infected node density for different initial infection densities, $\rho(0)$, in Fig.~\ref{fig:4}(a).
When $\beta_2 = 0$, the system exhibits no emotional group interaction effect, corresponding to the behavior of a standard SIS model.
However, as $\beta_2$ increases to $0.05$, a hysteresis loop emerges, indicating the presence of a bistable region, marked by the circular area between two dashed lines.
Within this region, two stable states coexist: $\rho^* > 0$ and $\rho^* = 0$, corresponding to a specific range of $\lambda_1$.
In contrast, this phenomenon does not occur when the value of $\beta_2$ is small, suggesting that the occurrence of the hysteresis loop depends on the strength of the group interaction effect.
This observation is consistent with the mean field analysis results shown in Fig.~\ref{fig:enter-label}(b). 

To further validate this observation, we track the evolution of the contagion dynamics, with the results presented in Figs.~\ref{fig:4}(b) and (c).
The color bar, transitioning from dark to light, represents the values of $\rho(0)$, ranging from low to high.
We uniformly sample 20 different initial densities of infected nodes within the range [0.01, 0.99]. 
The key distinction between panels (b) and (c) lies in their respective $\lambda_1$ values---one is selected within the bistable region, while the other lies outside of it.
Panel (b) shows convergence into two stable state, demonstrating the bistable phenomenon: one state remains near zero, while the other stabilizes around 0.6. 
In contrast, panel (c) shows convergence towards a single stable state, regardless of the initial infection density, $\rho(0)$.
The observation suggests that the bistable region  emerges only within a specific range of $\lambda_1$.

Figure~\ref{fig:4}(a)–(c) highlight the presence of pronounced hysteresis loops and bistable states when $q$ is low and the group interaction strength is set to $\beta_2 = 0.05$.
However, even with a fixed group interaction strength, the proportion of negative edges influences the existence of the bistable region.
To explore this, we examine the relationship between $q$ and the parameter range in which the bistable region exists.
We identify the critical thresholds $\lambda_c^{\uparrow}$ and $\lambda_c^{\downarrow}$ from the transitions observed in contagion processes initiated with different initial infection densities, based on the numerical simulation~\cite{radicchi2015predicting}. 
Specifically, $\lambda_c^{\uparrow}$ and $\lambda_c^{\downarrow}$ represent the thresholds for contagion starting with $\rho(0) = 0.01$ and $\rho(0) = 0.15$, respectively.
We then plot the relationship between $\lambda_c^{\uparrow}$, $\lambda_c^{\downarrow}$, and $q$ in Fig.~\ref{fig:4}(d) (circles and squares).

The results show that as $q$ increases, the gap between $\lambda_c^{\uparrow}$ and $\lambda_c^{\downarrow}$ gradually narrows,  indicating that the bistable region diminishes with a higher proportion of negative edges. 
This aligns with our mean field findings, where both hysteresis loops and bistable behavior fade as $\lambda_2$ decreases and approaches 1 shown in Fig.~\ref{fig:enter-label}(b).
This occurs because a higher proportion of negative edges weakens the emotional group interaction strength, as illustrated in Fig.~\ref{fig:3}(d).
Moreover, $\lambda_c^{\uparrow}$ and $\lambda_c^{\downarrow}$ eventually converge into a single line at the critical threshold $q_c$.
This point marks the transition from a discontinuous to a continuous phase transition, where the bistable phenomenon disappears. 
Beyond $q_c$, given a fixed group interaction strength, the system shifts to a monostable state with only one stable solution, indicating that the contagion process now follows a continuous phase transition with no bistable region remaining.

We notice that the numerically identified $\lambda_c^{\uparrow}$ exhibits a strong dependence on $q$, which contrasts with the mean field analytical result, where $\lambda_1^c = 1$ remains constant regardless of $q$.
Meanwhile, although the numerically identified $\lambda_c^{\downarrow}$ differs slightly from the mean field analytical result $\lambda_c$ given by Eq.~\ref{Eqs:lambdac}, the qualitative behavior remains consistent.
The discrepancies may arise from the limitations of the mean field approach, as the synthetic networks used in our numerical simulations exhibit higher-order correlations between group interactions that are not captured by the mean field theory. 
This difference further highlights that the proportion of negative edges can significantly shift the position of the critical point in contagion dynamics.

\section{Conclusion}
This study investigates the contagion dynamics in signed networks, integrating emotional group interactions and pairwise interactions, with a focus on how negative relationships shape the system’s dynamical behavior.
Through a combination of mean field analysis and numerical simulations, we derive several key insights.
First, we demonstrate that emotional group interactions can induce discontinuous phase transitions, resulting in bistable behavior and hysteresis loops under certain conditions. 
However, as the proportion of negative edges $q$ increases, the influence of group interactions weakens, altering both the nature and position of the phase transition.
When $q$ exceeds a critical threshold $q_c$, the system shifts to a monostable state, where only continuous phase transitions occur, and bistable regions vanish.
Our analysis highlights that the weakening of emotional group interactions, driven by increasing negative edges, as the primary factor behind these shifts in contagion dynamics.
Notably, the relationship between emotional group interaction strength and the proportion of negative edges exhibits nonlinear dependency, leading to uneven reductions in group interaction effects.
This reduction is slower at higher proportions of negative edges, emphasizing the distinct responses of pairwise and group interactions to changes in negative edge proportions, which leads to a shift in the dominant mechanisms governing contagion dynamics under different conditions.
These findings offer deeper insights into how negative edges shape phase transitions and the spread of contagion in signed networks.

In conclusion, this study sheds light on the critical role of emotional interactions in driving complex contagion dynamics within signed networks.
The interplay between pairwise and group interactions reveal the nontrivial impact of negative relationship on the dynamics behavior.
Understanding these dynamics is crucial for designing targeted interventions in social networks to mitigate the spread of harmful behaviors, such as misinformation, panic, or social unrest.
These insights offer new perspectives for managing complex contagion processes across diverse social contexts.

However, existing theoretical frameworks are still limited in capturing the full complexities of real-world scenarios. 
Future research should aim to refine these frameworks and assess the applicability of our model to observed phenomena. 
Moreover, further studies could explore the universality of the phase transition behaviors identified here across various empirical signed networks, providing a deeper understanding of contagion dynamics in complex social systems.

\appendix
\setcounter{figure}{0}
\renewcommand{\thefigure}{S\arabic{figure}}

\begin{figure}
    \centering
    \includegraphics[width=1\linewidth]{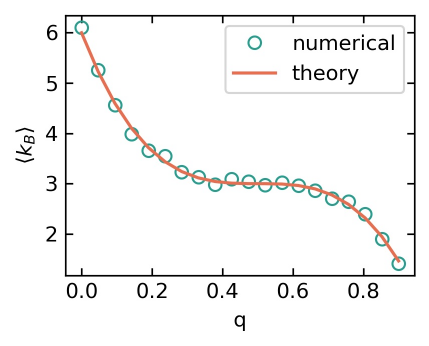}
    \caption{The numerical and theoretical results showing the relationship between \textit{q} and the average number of balanced triangles in a signed network. 
    The line represents the theoretical values, while the dots correspond to the numerical results.
    }
    \label{fig：numerical_theory}
\end{figure}

\end{document}